\documentclass[12pt]{iopart}
\usepackage{epsfig}
\usepackage{graphicx}
\usepackage{amssymb}

\newcommand{\here}{\makebox(0,0)}

\newcommand{\be}{\begin{equation}}
\newcommand{\ee}{\end{equation}}
\newcommand{\bd}{\begin{displaymath}}
\newcommand{\ed}{\end{displaymath}}

\newcommand{\bsigma}{{\mbox{\boldmath $\sigma$}}}

\newcommand{\bJ}{\ensuremath{\mathbf{J}}}

\newcommand{\bxi}{{\mbox{\boldmath $\xi$}}}

\newcommand{\sign}{{\mbox{sign}}}
\renewcommand{\to}{{\rightarrow}}

\begin{document}

\title[Survey propagation for the cascading Sourlas code]{Survey propagation for the cascading Sourlas code}

\author{ J P L Hatchett$^\dag$ and Y Kabashima$^\S$}
\address{\dag~ 96 Highbury Hill, London N5 1AT, UK}
\address{\S~ Department of Computational Intelligence and Systems
  Science, Tokyo Institute of Technology, Yokohama 226 8502, Japan}
\begin{abstract}
We investigate how insights from statistical physics, namely survey
propagation, can improve decoding of a particular class of sparse
error correcting codes. We show that a recently proposed algorithm,
time averaged belief propagation, is in fact intimately linked to a
specific survey propagation for which Parisi's replica symmetry
breaking parameter is set to zero, and that the latter is always
superior to belief propagation in the high connectivity limit. We
briefly look at further improvements available by going to the
second level of replica symmetry breaking.
\end{abstract}

 \ead{\tt hatchettman@hotmail.com,
kaba@dis.titech.ac.jp}
\section{Introduction}
Error correcting codes are an important device for communicating
information through noisy channels. We are living in what some
social commentators have called the ``information age'', with
apparently ever increasing appetite for broadband internet, mobile
phone data transmission, satellite television, computational power
etc. Hence, it is no surprise that such codes have are of
significant practical use.

A family of error-correcting codes based on an identification with
the ever popular Ising model from statistical mechanics were
proposed by Sourlas \cite{Sourlas89}, and novel coding methods were
examined \cite{Sourlas94, Rujan94}. A further important advance, in
terms of practical significance, was the move to finite rate, finite
length Sourlas type codes \cite{KabashimaSaad99} - which used
interactions between $K$ bits and $C$ interactions per bit. While
these codes showed good theoretical performance, the number of
successfully transmitted bits was significantly higher (for certain
channel noises) when $K > 2$. However, in this case the basin of
attraction (BOA) for a local decoding algorithm is very small
meaning that decoding is computationally extremely challenging for
reasonable code lengths. A more directly useful version of the above
theory involves cascading Sourlas codes \cite{KanterSaad00}. The
basic idea is to use multiple values of $K$ in the interactions. For
$K= 2$ (i.e. 2-body interactions) there is an energy landscape
dominated by the state which overlaps the sent message. Once an
algorithm has found this state with good overlap, higher body
interactions can be invoked which increase this overlap
substantially provided that the state is within the BOA of the high
overlap state.

Belief propagation (BP) \cite {Pearl88}, which is closely related
to the Bethe approximation and the cavity (replica) method in
finite connectivity disordered systems \cite{Yedidia,
 KabashimaSaad98}, has been the principle algorithm
used for decoding the finite rate Sourlas codes including the case
of the cascade version. Sometimes the quenched disorder in this
problem, in the form of corrupted bits which cause frustration in
the graph of bit interactions, brings about a certain disruption to
the BP dynamics, which can lead to deterioration of error correction
performance. Recently, a finite temperature survey propagation
algorithm (SP) \cite{Mezardetal02, MezardParisi01, MezardParisi03,
MezardZecchina02} was applied to a biased 3-body Sourlas code
\cite{Wemmenhove05}. The latter found that this more advanced
algorithm did lead to a shift in the critical channel noise for
computationally feasible decoding when BP did not converge. They
also noted that averaging the beliefs in time, or time averaged
belief propagation (TABP) \cite{vanMourik06,Bounkongetal05} could
improve decoding in such cases as well. However, the reason for this
improvement has not been fully clarified yet.

The objective of this article is to pinpoint the major contributing
factors for such improvement in practical decoding. For this, we
will investigate how the overlap between the original and estimated
messages in the first state of the cascading codes, for which only
$K=2$ interactions are considered, are influenced by the employed
decoding algorithms when the Almeida-Thouless (AT) stability is
broken \cite{deAlmeidaThouless78}. For simplicity of numerical
experiments, we will mainly focus on the zero temperature cases.
Although the model that we will deal with does not directly accord
with those in the preceding work \cite{Wemmenhove05}, the two
systems have a common feature that the AT stability is broken and,
therefore, insights gained from our simpler system is probably
useful for understanding what occurs in the biased 3-body
interaction system at finite temperature. By applying SP to the
first stage of a cascading Sourlas code, we will find that we can
increase the overlap with the original message below the
AT-transition, which in turn improves the chance of being within the
BOA for good decoding once we introduce higher body interactions. It
has also recently been shown that the first step of replica symmetry
breaking (1RSB) approach leads to higher values of the magnetisation
for disordered ferromagnetically biased $\pm J$ 2-spin Ising systems
on finite connectivity random graphs \cite{Castellani05}. Thus, we
strongly speculate that the higher first stage overlap gained by
introduction of the RSB ansatz in the cascading scheme is the main
factor for the improvement of the error correcting performance. This
speculation also suggests that algorithms based on increasing levels
of replica symmetry breaking (RSB) would improve matters further,
albeit at high computational cost. We make an initial investigation
of this by a brief analysis of a 2RSB survey propagation (2SP:
although this could rightly be called a survey of surveys
propagation). We will also argue that TABP corresponds to survey
propagation of a specific type for which the Parisi parameter, which
specifies the size of subgroups of replicas in the 1RSB ansatz, is
set to zero and is never bettered by the simple BP algorithm, as is
observed in CDMA multiuser detection problems
\cite{Kabashima05c,Kabashima05}.

This article is organized as follows. In the next section, we
introduce the model that we will study. In sections 3 and 4, two
decoding schemes based on BP and SP are presented in the case of
vanishing temperature. In sections 5 and 6, we will show numerically
that the SP based algorithm with zero Parisi parameter (SP0) is, in
practice, equivalent to TABP and provides a better performance than
BP when the AT stability is broken. This relationship of SP0 to
other algorithms is provided analytically in the vanishing code rate
limit in appendix A. The general proof of this relationship,
unfortunately, not been managed yet. We also show numerically  that
2SP can lead to an improvement of the decoding performance in the
situation where the AT stability broken. This implies that a further
improvement can be expected by introducing a higher level RSB ansatz
although the computational cost for decoding will be sacrificed. The
final section is devoted to a summary.

\section{Model definitions}
We assume that we have some message source vector $\bxi =
\{\xi_1,\ldots,\xi_N\} \in \{-1,1\}^N$ (for simplicity of exposition
we work within the Ising formalism). We encode this source vector
via a transformation to an $M$-bit vector $\bJ^0$ where $J^0_a =
\prod_{j \in a} \xi_j$ and where we have $M > N$. The rate $R$ of
such a code is given by $N/M$. For each set $a$, given its size, we
populate it with indices at random from $\{1,\ldots, N\}$ with each
index appearing exactly once in any given set $a$, and each index
appearing $C_{K}$ times in each $K$ body interaction. We take the
channel to be a binary symmetric channel so that when the encoded
message vector $\bJ^0$ is passed through the channel each bit is
flipped independently with probability $1-p$ and we denote the
received message by $\bJ$. The object of any decoding algorithm is
to estimate $\bxi$ with some $\bsigma$ using information from $\bJ$.
A natural performance measure is the overlap between our estimate
and the true message $M = \frac1N \sum_i \xi_i \sigma_i$. One way to
proceed with decoding is to look for the ground state of the
following Hamiltonian:
\begin{eqnarray}
\mathcal{H}(\bsigma) = - \sum_{a = 1}^M J_a \prod_{i \in a} \sigma_i
\label{eq:Hamiltonian}
\end{eqnarray}
Since this Hamiltonian is invariant under the transformation
$\sigma_i \to \sigma_i \xi_i$ and $J_{a} \to J_{a} \prod_{i \in a}
\xi_i$ we can gauge away the original message and are left with a
multispin ferromagnetically biased $\pm J$ spin glass model.

For the cascading approach we define a new auxiliary Hamiltonian
$\mathcal{H}^\prime(\bsigma) = -\sum_{a = 1}^M J_a^\prime \prod_{i
\in
  a}\sigma_i$ where $J_a^\prime = J_{a}$ if $|a| = 2$ and $J_a^\prime
= 0$ otherwise. The initial step is to find minima of
$\mathcal{H}^\prime(\bsigma)$ which is relatively easier than finding
minima of $\mathcal{H}$. We then use this minima as an initial
condition for local search algorithms looking for minima of the full
Hamiltonian (\ref{eq:Hamiltonian}).

\section{Solution at the level of replica symmetry}

An approximate method of finding the groundstate is available via
belief propagation, which is equivalent to the cavity method at the
level of replica symmetry for locally tree like graphs, which the
present code is in the long code length limit. The same
results can be found in the ensemble of graphs in the thermodynamic
(long code length) limit by use of the replica method, however, since
we are interested in decoding particular examples the former method is
preferable. The method has been discussed a great deal in recent
literature \cite{Pearl88, Yedidia, KabashimaSaad98,MezardParisi03} so
we will be relatively brief. We define cavity fields $h_{i \to a}$ to be
the field on site $i$ (decoded estimator $\sigma_i$) in the absence of
link $a$ (received bit $J_a$)  and the message $u_{a
  \to i}$ to be the message from received bit $J_a$ to decoded bit
$\sigma_i$. We are working with a discrete energy model and a
self-consistent solution to these equations is given by restricting
the values of $\{u, h\}$ to integers. Then,
the belief propagation (BP)
update equations
describing these variables are:
\begin{eqnarray}
h_{i \to a}^{t+1} = \sum_{b \in \mathcal{N}(i)\setminus a}
u_{b \to i}^t,
\label{eq:cavityupdates_rs0}\\
u_{a \to i}^{t} = \sign(J_{a} \prod_{j \in a \setminus i} h_{j \to
  a}^t) \label{eq:cavityupdates_rs}
\end{eqnarray}
where $\mathcal{N}(i)$ denotes the set $\{a : i \in a\}$ and we take
the convention $\sign(0) = 0$. Other than
the trivial paramagnetic state ($h_{i \to a} = u_{a \to i} = 0 \
\forall i,a$), fixed points
of the message passing equations (\ref{eq:cavityupdates_rs})
correspond to local minima of the energy of
the Hamiltonian (\ref{eq:Hamiltonian}). However, it is certainly not
guaranteed that the equations (\ref{eq:cavityupdates_rs}) will
converge at all. If the messages converge, or after some specified number
of updates if they do not, the estimator for the original message bits
is given by
\begin{eqnarray}
\sigma_i = \sign(\sum_{a \in \mathcal{N}(i)} u_{a \to i})
\end{eqnarray}
Consider a specific example, that of a code with $C_2 = 3$ and $C_3
= 3$ so each original message bit $\xi_i$ is contained within three
2-body interactions and three 3-body interactions. Due to the 3-body
interactions we find that without using the cascading approach we
never find the state with high overlap - even at zero channel noise
- due to the higher body interactions, which is similar to the
result found for purely ferromagnetic 3-body interactions in
\cite{Franzetal02}, where even for these purely ferromagnetic
interactions the structure of the interactions led to glassiness.
However, if we first allow the system to equilibrate using the
energy landscape given by only the 2-body interactions (so perform
BP on the auxiliary Hamiltonian) we find that for a certain range of
$p$ we obtain non-zero values of $M$. Since for purely 2-body
interactions there is a reflection symmetry in the Hamiltonian there
is no bias at this stage between $\pm M$. However, when the 3-body
interactions are added to the Hamiltonian it is possible to infer
which of these two states gives a better overlap overall, namely the
state with lower energy. BP is then performed on the Hamiltonian
with the 3-body interactions added. In practice, we add the the
3-body interactions in three batches, at each batch we add one
3-body interaction to the Hamiltonian for each spin (so $N/3$
interactions). Since the initial state for the combined system is
one with finite $M$, it is closer to the ground state with high
overlap than a random initial condition so BP has a better chance of
reaching the ground state.

\section{Solution at the level of replica symmetry breaking}

Past some critical channel noise we expect there to be an ergodicity
breaking transition corresponding to replica symmetry breaking, at
which point the update equations (\ref{eq:cavityupdates_rs}) will no
longer be sufficient to correctly characterise the system. For
example, if we consider the initial stage of the algorithm where
messages are passed purely between sites and 2-body links then it is
known that for $C_2 = 3$ (three 2-body interactions) the
AT-transition occurs at $p = \frac{11}{12}$ \cite{KwonThouless88,
Castellani05}. Below this point a survey propagation algorithm
\cite{Mezardetal02, MezardZecchina02} should improve at least the
initial overlap $M$ (the AT-point for the full system will not be
the same as that for the initial system) - at most until the
transition from the mixed phase (glassy ferromagnetic pure states)
to the spin glass phase. By considering the finite temperature
replica symmetric transition lines away from the paramagnetic phase,
a bifurcation analysis of the belief propagation equations yields
critical temperatures for the $C_2 = 3$ regular Sourlas code:
\begin{eqnarray}
P \to F: \qquad &1 = 2 (2p-1) \tanh(\beta)\\
P \to SG: \qquad&1 = 2 \tanh^2(\beta)
\end{eqnarray}
which implies the triple point is at $p = \frac12(1 +
\sqrt{\frac12}) = 0.85355\ldots$. According to the Parisi-Toulouse
hypothesis of no re-entrance this gives us the zero temperature
transition point between the mixed phase and the spin glass phase.

\begin{figure}[t]
\vspace*{-2mm} \hspace*{45mm} \setlength{\unitlength}{0.75mm}
\begin{picture}(300,85)
\put(-45,15){\epsfysize=60\unitlength\epsfbox{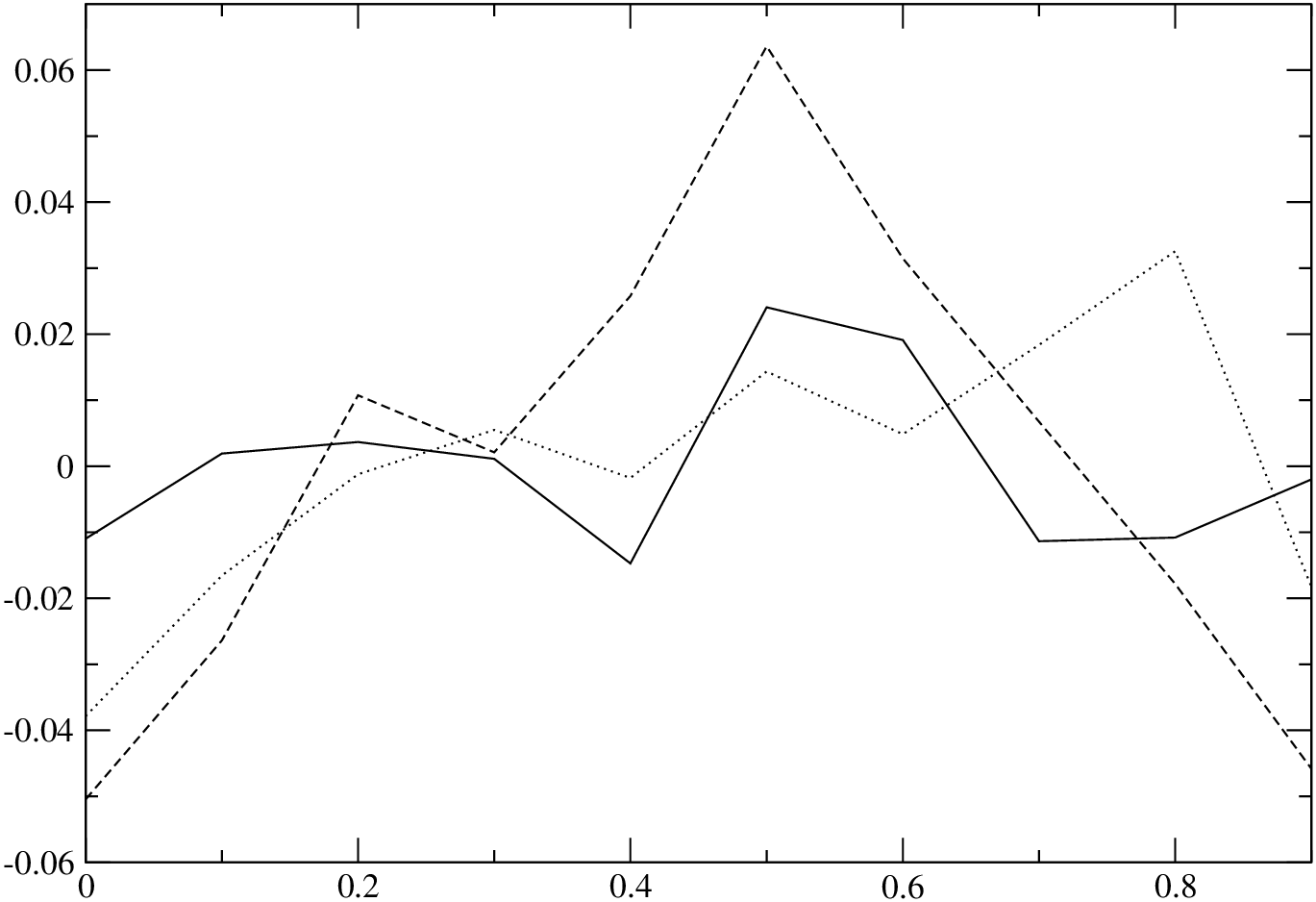}}
\put(0,6){\here{$x$}}
\put(55,15){\epsfysize=60\unitlength\epsfbox{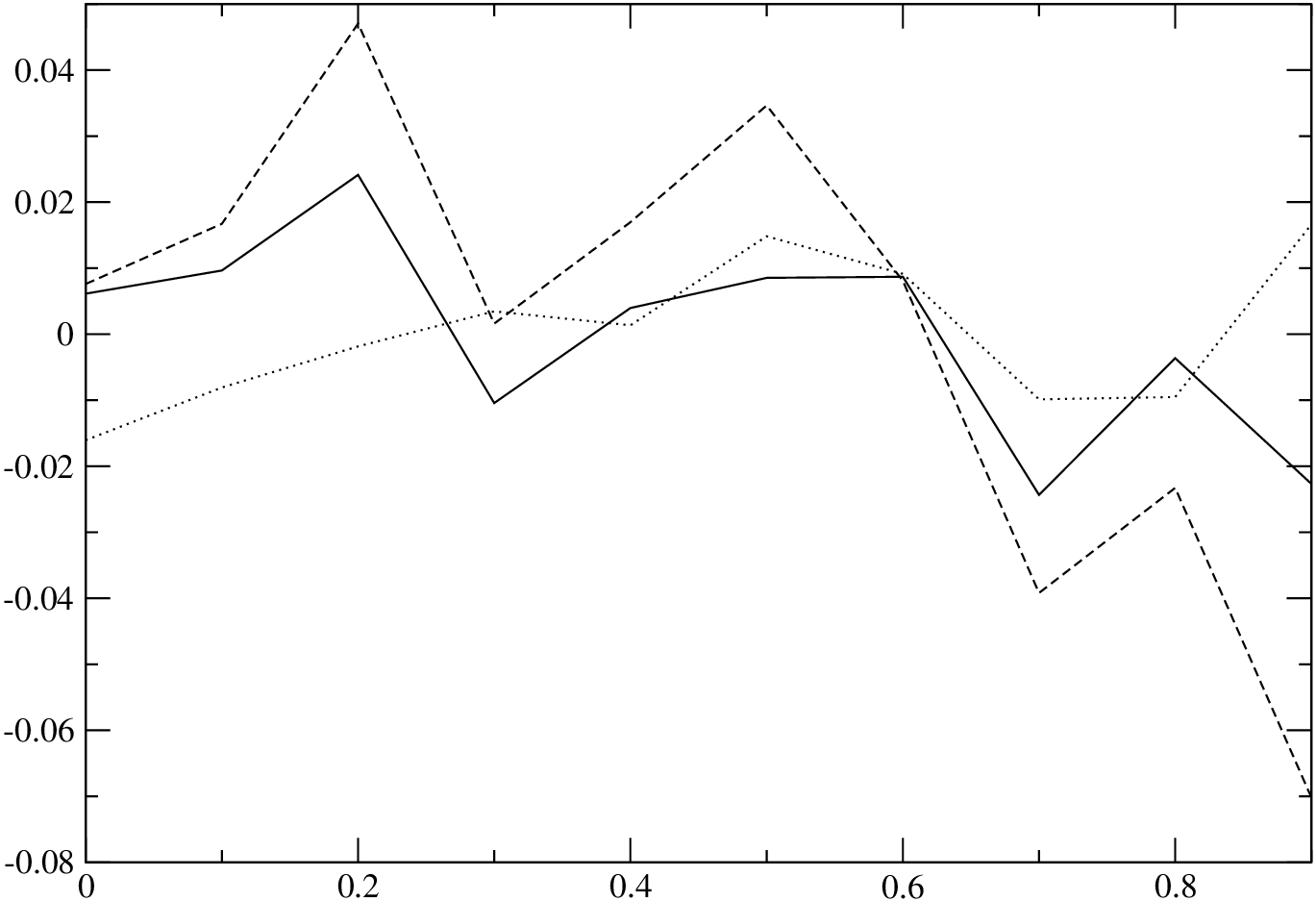}}
\put(95,6){\here{$x$}}
\end{picture}
\vspace*{-8mm}
\caption{Variation (from the mean) in the overlap (solid
line), energy
  (dashed line) and (zero-temperature)
  free energy (dotted line) versus the zero-temperature Parisi
  parameter x. The data are averages over 25 run with code lengths $N
  = 1000$. The left figure is for $p = 0.86$ while the right figure is
  for $p = 0.88$. In both figures we see that variation in the overlap
  follows variation in the energy to a reasonable extent.}
\label{fig:overlapvsx}
\end{figure}

At
the first level of replica symmetry breaking, survey propagation with
Parisi's replica symmetry breaking parameter equal to $x$ (SPx) rather
than solving for a single message from sites to links (and links to
sites) we have to solve for a distribution of messages. This
distribution is over the messages in different ergodic sectors. While
true ergodicity breaking will not occur in a finite length message,
the breaking effects are sufficiently strong to mean that it does
occur to all intents and purposes on any practical timescale for a
message of reasonable length. The key idea behind the SPx is that when
performing an iteration of our messages, the energies in different
ergodic sectors will vary, and as they vary the probability of finding
such a given state also varies. The SPx equations keep track of these
changes assuming that the true nature of the states is 1RSB (which is
not necessarily the case). If we again restrict ourselves to integer
values for the messages (so in fact the integrals are really sums over
the appropriate domain) then the iteration equations are:
\begin{eqnarray}
P_{i\to a}^{t+1}(h) \propto \int [\prod_{b \in \mathcal{N}(i) \setminus a}
  \rmd u_b Q_{b \to i}^t (u_b)] \delta[h - \sum_b u_b] \rme^{x
  \left|\sum_b u_b\right|}\\
Q_{a \to i}^t(u) \propto \int [\prod_{j \in a \setminus i} \rmd h_j P_{j
  \to a}^t(h_j)] \delta[u - \sign(J_{a} \prod_{j} h_j)] \rme^{x
  \delta_{u,0}} \label{eq:SPx}
\end{eqnarray}
The above pair of equations are iterated until convergence (or for
some sufficient number of iterations). At that point we can construct
our estimate for the decoded message via:
\begin{eqnarray}
\sigma_i = \sign(\int \rmd H P_i(H) \sign(H))
\label{MPMdecoder}\\
P_i(H) \propto \int [\prod_{b \in \mathcal{N}(i)}
  \rmd u_b Q_{b \to i} (u_b)] \delta[H - \sum_b u_b] \rme^{x
  \left|\sum_b u_b\right|}
\end{eqnarray}
The approach thus far has still left the choice of $x$. In physical
problems, the general approach is to extremeise the generalized free
energy wrt $x$ \cite{MezardParisi01} as these states are the ground
states of the system. However, despite using a physical approach, we
are interested in the error correcting abilities of this system and
it does not necessarily hold that extremising the zero temperature
free energy corresponds to maximising the overlap $M$. In fact, we
have found that minimizing the energy $\mathcal{H}(\bsigma)$ gives a
good guide to choice of $x$ which is reasonable since our initial
aim was to decode via minimizing the energy of the Hamiltonian
(\ref{eq:Hamiltonian}). If the computational effort required to fit
the model is prohibitive, choosing a particular value of $x$ for SPx
(we found about 0.15 was reasonable) still gives a marked
improvement over BP in decoding across a range of channel noises
$p$. In figure \ref{fig:overlapvsx} we plot variation in the overlap
from its mean value to that of the energy and the free energy (we
take 10 times the variation of the free energy so that the scales
are similar). The energy (which is observable) seems to give a
reasonable indication of the overlap (which is not observable in
practical situations).

It is to be expected that to correctly describe state space up to
the transition from a mixed (glassy ferromagnetic) phase to a spin
glass phase full RSB needs to be taken into account. However, this
is not feasible computationally. This does make the next obvious
step a 2RSB algorithm, 2SPxy, where the messages will be surveys of
surveys. For a more detailed discussion on applications to $k$-SAT
see \cite{Montanari04}. We restrict our discussion for simplicity to
the first stage of the decoding procedure (where all interactions
are 2-body). Then we have update equations;
\begin{eqnarray}
\mathcal{P}_{i \to a}^{t+1}[P] \propto \int \prod_{b \in \mathcal{N}(i)
  \setminus a} \rmd \mathcal{Q}_{b \to i}^{t}[Q_{b \to i}] z[\{Q\}; y]^{x/y}
  \delta[P - P[\{Q\}; y]]\\
\mathcal{Q}_{a \to i}^t[Q] = \int \rmd \mathcal{P}_{j \to a}^t[P] \delta[Q
  - Q[P;a]]\\
P[\{Q\};y]= \frac{1}{z[\{Q\};y]} \int \prod_{b \in \mathcal{N}(i) \setminus
a} \rmd u_b
  Q_b(u_b) \delta[h - \sum_b u_b] \rme^{y (|\sum_b u_b| - \sum_b |u_b|)}\\
Q[P;a](u) = \int \rmd h P(h) \delta[u - \sign(J_a h)]
\end{eqnarray}
where $z[\{Q\};y]$ is the normalising constant for $P[\{Q\};y]$ and
$x$ and $y$ are the zero temperature Parisi parameters. The
integrals over $u,h$ are sums over the appropriate integers while
the integrals over measures are over distributions on the integers.
We will compare these different approaches numerically in the
following.

\section{Time averaged belief propagation and SP0}
In general there is no guarantee that BP will converge on a graph
with loops. However, experimentally it appears that for large
disordered random graphs the transition to non-convergence is
closely linked to the AT-transition~\cite{Kabashima03}. One can also
rationalise this in terms of correlation lengths. If the correlation
length is less than log($N$) then locally the graph appears to be a
tree (apart from at a finite number of sites), so BP should converge
provided the graph is sufficiently large (of course this is very
large as the length scale is log($N$)). Below the AT transition due
to the proliferation of pure states there are necessarily
correlations on all length scales throughout the graph, so one would
not expect BP to converge as it is affected by the loops. One
simple, robust approach to deal with estimation when BP has not
converged is to use time averaged BP (TABP), which corresponds to
taking averages in time of the messages, and performing inference
with these average messages \cite{vanMourik06, Bounkongetal05,
Wemmenhove05}.

\begin{figure}[h]
\vspace*{-2mm} \hspace*{45mm} \setlength{\unitlength}{0.75mm}
\begin{picture}(200,105)
\put(0,10){\epsfysize=80\unitlength\epsfbox{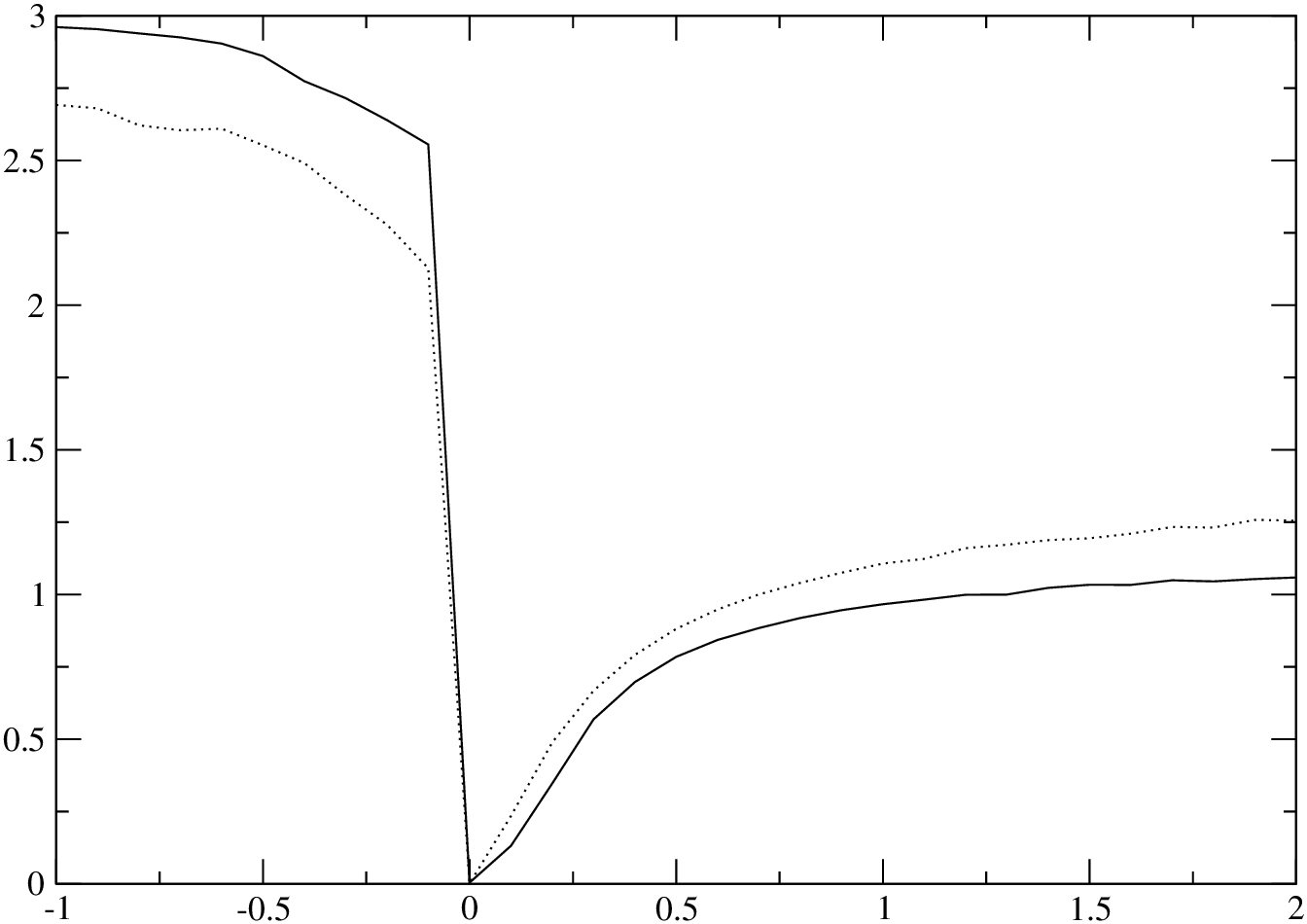}}
\put(57,6){\here{$x$}} \put(-10,50){\here{$D(x)$}}
\end{picture}
\vspace*{-8mm} \caption{We see how the average distance $D(x)$
between the message distributions from SPx and the message
distributions from TABP vary with x. The solid line is for channel
noise $p = 0.88$ while the dotted line is for $p = 0.86$. We find
good support for the hypothesis that TABP is equivalent to SP0.}
\label{fig:distances}
\end{figure}

Another algorithm which improves on BP but is simpler than full SPx
is SP0, as used for the CDMA problem in
\cite{Kabashima05c,Kabashima05}. SP0 does not resort to any
reweighing of messages as the messages are passed but gains its
improvement from a ``majority rules'' approach. I.e. when decoding
in a region where BP does not converge, inference is performed at
each bit by using the majority decision of the different instances.
For the fully connected CDMA problem, the local fields are Gaussian
and it is possible to show analytically that SP0 improves inference
compared to BP analytically~\cite{Kabashima05}. In the present
situation due to the finite connectivity nature of the problem the
effective fields are not Gaussian distributed, thus we have not been
able to show that SP0 improves on BP analytically (basically we do
not have sufficiently general control of $\{Q(u), P(h)\}$). However,
in the high $C$ limit our present model reduces to the SK-model
\cite{SherringtonKirkpatrick75} and in this case we again find that
SP0 is strictly better in inference terms than BP below the AT-line,
which is shown in appendix A.

In fact, TABP and SP0 are intuitively quite similar.
The first averages
over BP states in time, while the latter averages over different BP
states in parallel from different initial conditions and with mixing
of different solutions at each update step. Provided the dynamics of BP are
sufficiently chaotic that it explores all states one could expect both
averages to give similar (or even identical) results. To test this
hypothesis we compare the distribution of local fields given by SP0 to
the distribution given by TABP. More specifically we examine the order
parameter
\begin{eqnarray}
D(x) = \frac1N \sum_i \sum_{a \in N(i)} D_{H}[P_{i \to a}^{SPx} ||
P_{i \to a}^{TABP}]
\end{eqnarray}
with the Hellinger distance $D_{H}(P || Q) \equiv 2 \int \rmd x
(\sqrt{P(x)} - \sqrt{Q(x)})^2$. In, figure \ref{fig:distances} we look at
$D(x)$ vs
$x$ and see that we find very good agreement for the hypothesis that
SP0 is equivalent to TABP which gives an alternative explanation to
its improved performance over BP and provides an new method
for implementing SP0.

\section{Comparison of the different decoding algorithms}

\begin{figure}[h]
\vspace*{-2mm} \hspace*{45mm} \setlength{\unitlength}{0.75mm}
\begin{picture}(200,105)
\put(0,10){\epsfysize=80\unitlength\epsfbox{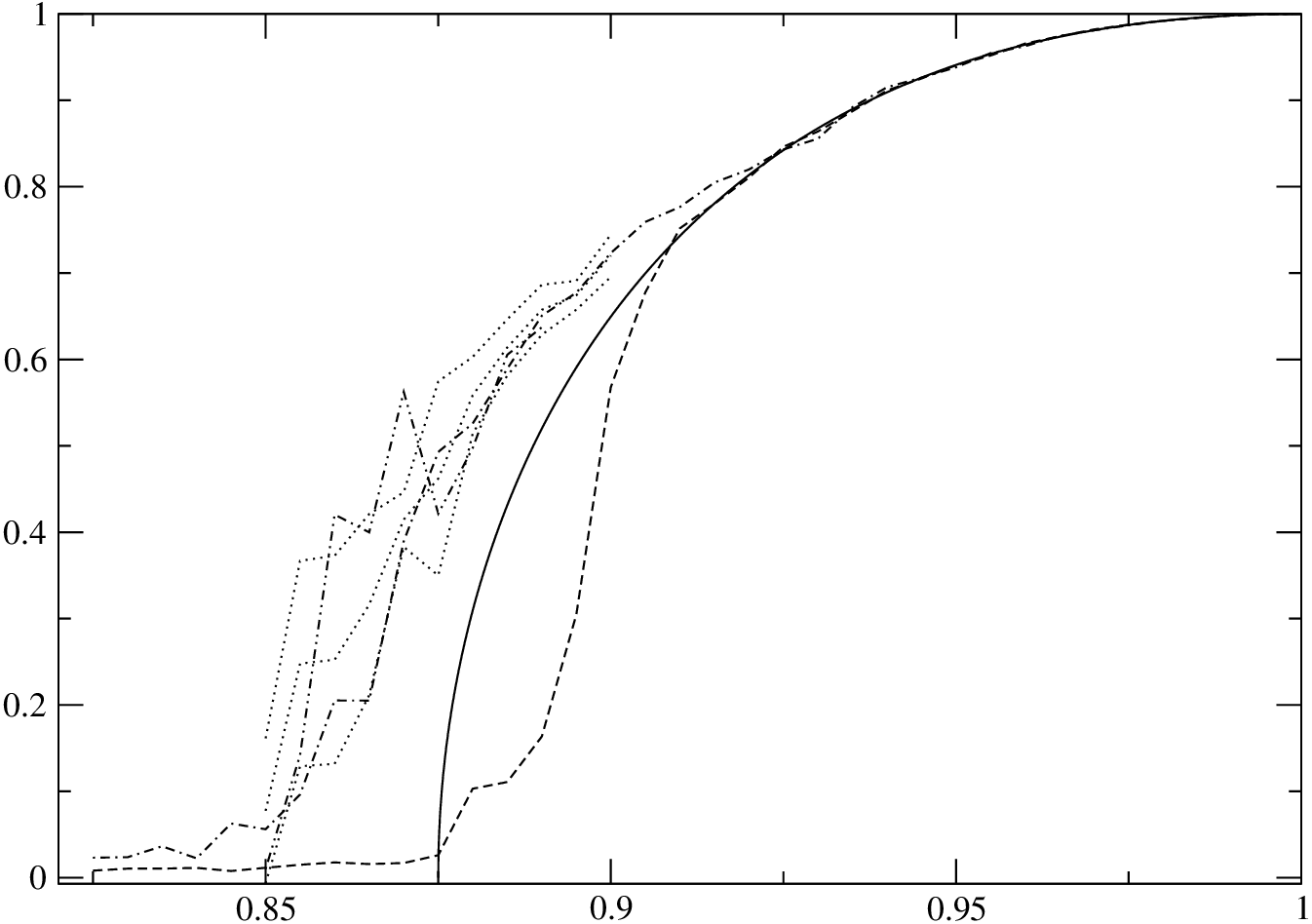}}
\put(57,6){\here{$p$}} \put(-10,50){\here{$M$}}
\end{picture}
\vspace*{-8mm} \caption{The value of the initial overlap $M$ versus
  the channel noise $p$ for several different algorithms. The solid
  line is the theoretical replica symmetric result in the infinite
  system sized limit. The dashed line is the average result (over 10
  runs) of BP (note agreement is good after $p \approx 0.915$ the
  theoretical AT point). The dot-dashed line is TABP (averages over 10
  runs) which improves
  significantly on BP for certain $p$. The three dashed lines are SPx
  (averaged over 10 runs and optimised over $x$)
  $\pm$ one standard deviation. The improvement of SPx on TABP is
  relatively modest. Finally the dot-dot-dashed line is a single run
  of 2SPxy with $x = 0.15$ and $y = 0.1$ (so not optimised) - it
  appears to improve on SPx for $p = 0.86,0.865, 0.87$. Note that the
  predicted transition from a mixed phase to spin-glass phase is at $p
  \approx 0.85355$.}
\label{fig:initial_mag}
\end{figure}

We have used four different algorithms for the initial decoding, BP,
TABP (or SP0), SPx and 2SPxy, corresponding to different levels of
replica symmetry breaking and model fitting. For each algorithm
there is essentially only one pertinent question which is whether
the overlap reached after initial decoding is sufficient to lead to
good full decoding. While it is tempting to think of some critical
$M$ - in reality this is not sufficient to describe the state in
full detail and other factors such as $p$ and the decoding
algorithms themselves will all contribute to changes in the $M$
required (in addition to finite size fluctuations) to reach the high
overlap state. However, due to issues of computational complexity we
have tested all four algorithms on the initial decoding only. The
2SPxy algorithm has only been investigated very briefly there, for a
single value of $x$ and $y$ rather than optimising over them for
each codeword received. We compare the results of these initial
overlaps in figure \ref{fig:initial_mag}. We see that each
improvement in the algorithm does lead to an increased overlap after
the initial stage, between the beginning of the mixed phase and the
AT point. The 2SPxy does appear (as far as we can tell from the
data) to give an improvement on the SPx algorithm for a limited
range of channel noise.

\begin{figure}[t]
\vspace*{-2mm} \hspace*{45mm} \setlength{\unitlength}{0.75mm}
\begin{picture}(200,105)
\put(0,10){\epsfysize=80\unitlength\epsfbox{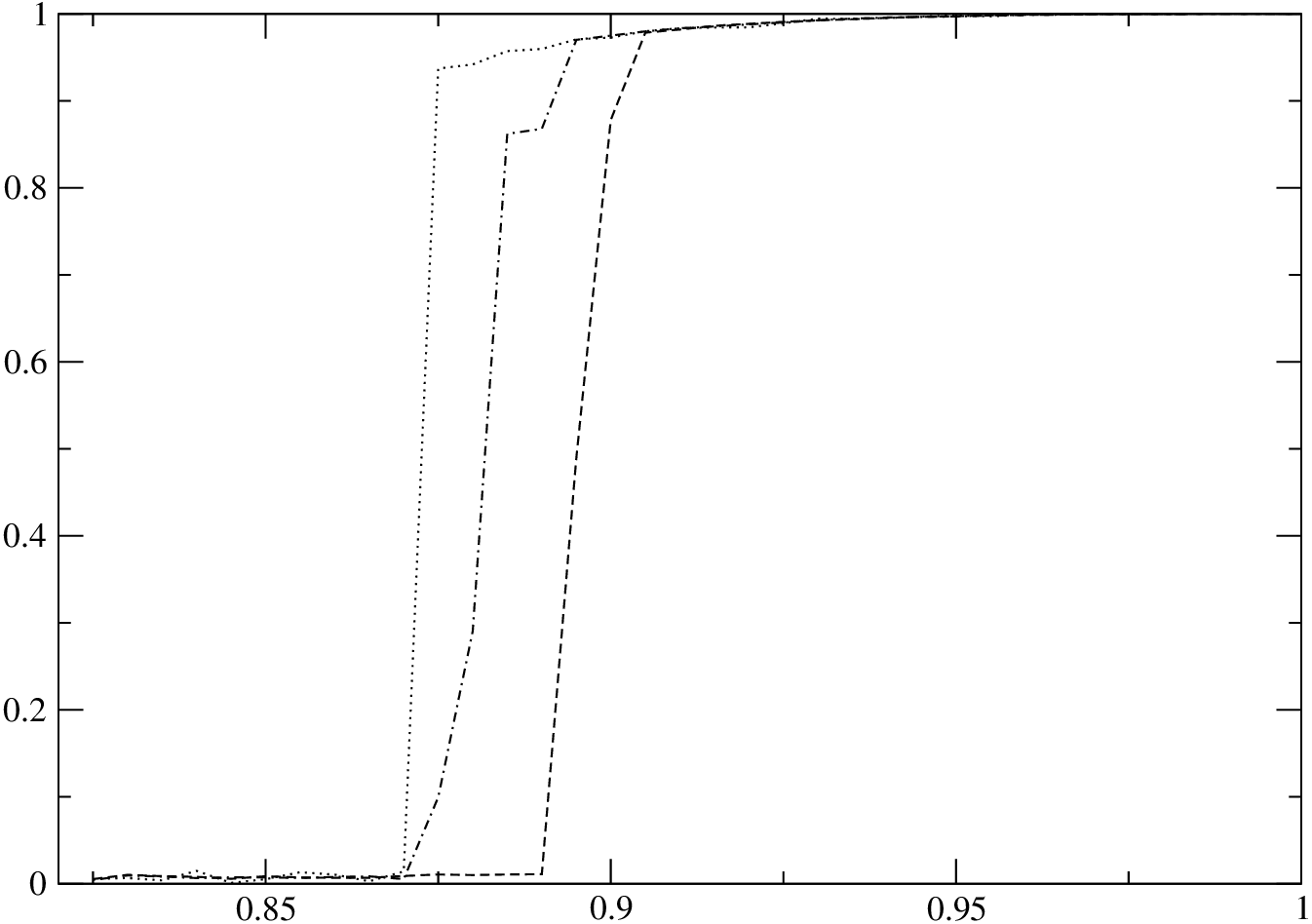}}
\put(57,6){\here{$p$}} \put(-10,50){\here{$M$}}
\end{picture}
\vspace*{-8mm} \caption{The final overlap $M$ versus $p$ averaged over
10 runs of code length $N = 10\,000$ for $BP$ (dashed line), TABP
(dot-dashed line) and SPx (dotted line). Each advance in algorithm
leads to an improvement in the critical channel noise. }
\label{fig:final_mag}
\end{figure}

Next we look in figure \ref{fig:final_mag} at the final overlap
reached after full decoding with BP, SP0 and SPx. We see that
improving the quality of the algorithm leads to a marked improvement
in the critical channel noise enlarging the space for which the
cascading code approach will work.

\section{Conclusion}

We have examined how taking replica symmetry breaking into account can
improve decoding of cascading Sourlas codes. The basis behind our
idea is that below the AT transition, the true overlap at the minima
of the Hamiltonian (\ref{eq:Hamiltonian}) is above its
replica symmetric estimate so increasing steps of replica symmetry
breaking will serve to improve the overlap between the estimate of the
message and the true message. We have also shown that TABP is closely
related (if not identical) to SP0 - this throws up two obvious areas
for further investigation, one to find under what conditions these
algorithms are identical and secondly to show that both of them are
strict improvements on BP below the AT transition generally. Another
area to improve is in terms of the 2SP, where further optimisation of
the Parisi parameters should lead to further improvements, albeit at
some considerable computational cost. The graph structure used here is
not expected to be optimal - after the initial stage using higher body
interactions (i.e. $K > 3$)  should give improvements in the final overlap,
optimising
over possible architectures should also improve matters, but the
principles developed here will help in any case where the channel
noise is sufficiently high that decoding is taking place below the AT
transition.

\ack
JPLH would like to thank Tommaso Castellani and Bastian Wemmenhove
for helpful discussion of their work. YK acknowledges support by
Grants-in-Aid Nos.~14084205 and 17340116 from MEXT/JSPS, Japan.

\appendix
\section{Superiority of SP0 to BP in dense connectivity limit}
Here, we show analytically that SP0 is never bettered by BP in terms
of the error correction performance when the number of connections
tends to infinity. Such a property is likely to hold in the case of
sparse connectivity as well, analytical proof of which has not yet
been managed.

\subsection{SPx for densely connected systems}
For simplicity and generality, we focus on regular $K=2$ systems at
general {\em finite} temperature $T=\beta^{-1}$. Extension to other
$K$ values and the case of vanishing temperature is straightforward.
SPx of this system can be expressed as
\begin{eqnarray}
P_{i \to a}^{t+1}(h)
&\propto &
\int [
\prod_{b \in {\cal N}(i) \backslash a}
\frac{du_b Q_{b \to i}^t(u_b)}{
(2 \cosh (\beta u_b))^{\frac{x}{\beta}}}
] \delta [h - \sum_b u_b]
(2 \cosh (\beta \sum_b u_b))^{\frac{x}{\beta}},
\label{SPx_SK1}
\\
Q_{a \to i}^t(u) &\propto & \int [ \prod_{j \in a \backslash i} d
h_j P_{j \to a}^t(h_j)] \delta [ u-u(h_j,J_a;\beta)],
\label{SPx_SK2}
\end{eqnarray}
where $u(h_j,J_a;\beta)\equiv \beta^{-1}\tanh^{-1}(\tanh(\beta J_a)
\tanh(\beta h_j) )$. In order to obtain a simpler expression for
$C_2=C \to \infty$, we introduce appropriate scalings $|J_a| \to
\frac{J}{\sqrt{C}}$ and $p=\frac{1}{2}+\frac{J_0}{2J \sqrt{C}}$,
where $J_0>0$ and $J>0$. This corresponds to situations of low code
rate communication that copes with strong noise despite restricted
signal power. Let us denote $\int dh_i P_{i \to a}^t(h_i)
\tanh(\beta h_i)= m_{i \to a}^t$, $\int dh_i P_{i \to a}^t(h_i)
\tanh^2(\beta h_i)= M_{i \to a}^t$ and characterize the system
utilizing two macroscopic variables $C^{-1}\sum_{b \in {\cal N}(i)
\backslash a } M_{i \to b}^t \sim C^{-1}\sum_{a \in {\cal N}(i) }
M_{i \to a}^t \sim (NC)^{-1} \sum_{i,a \in {\cal N}(i)}M_{i \to a}^t
\equiv Q_1^t$ and $C^{-1}\sum_{b \in {\cal N}(i) \backslash a }
(m_{i \to b}^t)^2 \sim C^{-1}\sum_{a \in {\cal N}(i) } (m_{i \to
a}^t)^2 \sim (NC)^{-1} \sum_{i,a \in {\cal N}(i)}(m_{i \to a}^t)^2
\equiv Q_0^t$ based on the law of large numbers. These, in
conjunction with the central limit theorem, indicate that eq.
(\ref{SPx_SK1}) is reduced to
\begin{eqnarray}
P_{i \to a}^{t+1}(h) =
\frac{(2\cosh (\beta h))^{\frac{x}{\beta}}
e^{-\frac{(h-\phi_{i \to a}^{t+1})^2}{2\Delta^{t+1}}}}
{\int dh
(2\cosh (\beta h))^{\frac{x}{\beta}}
e^{-\frac{(h-\phi_{i \to a}^{t+1})^2}{2\Delta^{t+1}}}},
\label{SPx_reduced1}
\end{eqnarray}
where
\begin{eqnarray}
\phi_{i \to a}^{t+1}&=&\sum_{b \in {\cal N}(t)\backslash a}
J_b m_{j \to b}^{t},\label{SPx_SK_simple1}\\
\Delta^{t+1} &=& \sum_{b \in {\cal N}(t)\backslash a}
J_b^2 (M_{j \to b}^t -(m_{j \to b}^{t})^2)
= J^2(Q_1^t-Q_0^t).
\label{SPx_SK_simple2}
\end{eqnarray}
In turn, eq. (\ref{SPx_reduced1}) provides expressions of
microscopic variables
\begin{eqnarray}
m_{i \to a}^t&=&\frac{\int dh (2 \cosh (\beta h))^{\frac{x}{\beta}}
e^{-\frac{(h-\phi_{i \to a}^{t})^2}{2\Delta^{t}}}
\tanh(\beta h)}{
\int dh (2 \cosh (\beta h))^{\frac{x}{\beta}}
e^{-\frac{(h-\phi_{i \to a}^{t})^2}{2\Delta^{t}}}},
\label{SPx_reduced2_1}\\
M_{i \to a}^t&=&\frac{\int dh (2 \cosh (\beta h))^{\frac{x}{\beta}}
e^{-\frac{(h-\phi_{i \to a}^{t})^2}{2\Delta^{t}}}
\tanh^2(\beta h)}{
\int dh (2 \cosh (\beta h))^{\frac{x}{\beta}}
e^{-\frac{(h-\phi_{i \to a}^{t})^2}{2\Delta^{t}}}}.
\label{SPx_reduced2_2}
\end{eqnarray}
Eqs. (\ref{SPx_SK_simple1})-(\ref{SPx_reduced2_2}) constitute the
simpler expression of SPx in the dense connectivity limit. Utilizing
$m_{i \to a}^t$ and $\Delta^t$, the estimate for the decoded message
at the $t$th update is provided as
\begin{eqnarray}
\sigma_i={\rm sign}(
\int dHP_i(H)\tanh(\beta H)), \label{MPM}\\
P_i^t(H)=\frac{(2\cosh (\beta H))^{\frac{x}{\beta}}
e^{-\frac{(H-\phi_i^t)^2}{2\Delta^t}}}{
\int dH (2\cosh (\beta H))^{\frac{x}{\beta}}
e^{-\frac{(H-\phi_i^t)^2}{2\Delta^t}}} ,
\label{localField}
\end{eqnarray}
where $\phi_i^{t}=\sum_{a \in {\cal N}(i)}J_a m_{j \to a}^{t-1}$.

\subsection{Macroscopic dynamics}
The performance of the SPx decoding algorithm
(\ref{SPx_SK_simple1})-(\ref{SPx_reduced2_2}) can be examined by
following time evolution of several relevant macroscopic
variables~\cite{Kabashima03,Kabashima03b,RichardsonUrbanke01}. The
maximum entropy principle, in conjunction of the law of large
numbers, makes it possible to regard the microscopic variable
$\phi_{i \to a}^{t}$ as Gaussian random numbers, the average and
variance of which are provided independently of site and check
indices $i$, $a$ as $E^{t}$ and $F^{t}$, respectively. We assume
that the gauge transformation $\sigma_i \to \sigma_i \xi_i$ and $J_a
\to J_a \prod_{i \in a}\xi_i$ is already carried out, which implies
that the original message is specified by $(1,1,\ldots,1)$. These
indicate that $Q_1^t$, $Q_0^t$ and the macroscopic overlap to the
original message $m^t=(NC)^{-1}\sum_{i,a \in {\cal N}(i)} m_{i \to
a}^t$ are represented by
\begin{eqnarray}
Q_1^t&=&\int  Dv
\frac{\int  Du (2\cosh
(\beta \Phi^t(u,v)))^{\frac{x}{\beta}}
\tanh^2(\beta \Phi^t(u,v))}{
\int  Du (2\cosh
\beta \Phi^t(u,v))^{\frac{x}{\beta}}},
\label{1RSB_Q1}\\
Q_0^t&=&\int  Dv
\left ( \frac{\int  Du (2\cosh  (\beta
\Phi^t(u,v)))^{\frac{x}{\beta}}
\tanh(\beta  \Phi^t( u,v))}{
\int  Du (2\cosh
(\beta \Phi^t(u,v)))^{\frac{x}{\beta}}} \right )^2,
\label{1RSB_Q0}\\
m^t&=&\int  Dv
\frac{\int  Du (2\cosh (\beta
\Phi^t(u,v)))^{\frac{x}{\beta}}
\tanh( \beta  \Phi^t(u,v ))}{
\int  Du
(2\cosh (\beta \Phi^t(u,v)))^{\frac{x}{\beta}}},
\label{1RSB_m}
\end{eqnarray}
respectively, where $Dz=\frac{dz e^{-z^2/2}}{\sqrt{2\pi}}$ and
$\Phi^t(u,v)=\sqrt{\Delta^t} u +\sqrt{F^t} v + E^t$. On the other
hand, the self-averaging
property~\cite{MezardParisiVirasoro87,Kabashima03} and eqs.
(\ref{1RSB_Q1}), (\ref{1RSB_Q0}) and (\ref{1RSB_m}) imply that
$F^{t+1}$ and $E^{t+1}$ are provided by
\begin{eqnarray}
F^{t+1}&=&N^{-1}\sum_{i,a \in {\cal N}(i)}
\left (\overline{(J_a m_{j \to a}^t)^2}
-(\overline{J_a m_{j \to a}^t})^2 \right )\cr
&\simeq &
N^{-1}\sum_{i,a \in {\cal N}(i)}\overline{(J_a m_{j \to a}^t)^2}
\simeq J^2 Q_0^t,
\label{1RSB_F}\\
E^{t+1} &=& N^{-1}\sum_{i,a \in {\cal N}(i)}
\overline{J_a m_{j \to a}^t}\simeq
J_0 m^t,
\label{1RSB_E}
\end{eqnarray}
respectively, where $\overline{\cdots}$ denotes the configuration
average with respect to the channel noise and lattice configuration.
The update rule for $\Delta^{t+1}$ is already provided by eq.
(\ref{SPx_SK_simple2}). Using these macroscopic variables, the error
probability per bit for SPx, $P_b(x)$, is calculated from
\begin{eqnarray}
P_b(x)&=&\int Dv \Theta\left (-
\frac{\int Du (2\cosh (\beta \Phi^t(u,v)))^{\frac{x}{\beta}}
\tanh( \beta  \Phi^t(u,v ))}{
\int  Du
(2\cosh (\beta \Phi^t(u,v)))^{\frac{x}{\beta}}}
\right ) \cr
&=&\int^{-\frac{E^t}{\sqrt{F^t}}}_{-\infty} Dv= {\rm erfc}\left
(\frac{E^t}{\sqrt{F^t}} \right ),
\label{bit_error_rate}
\end{eqnarray}
where $\Theta(z)=1$ and $0$ for $z>0$ and $z<0$, respectively.
It may be remarked  that
eqs. (\ref{SPx_SK_simple2}) and
(\ref{1RSB_Q1})-(\ref{1RSB_E}) accord
with the macroscopic description of the SPx dynamics for
the SK model~\cite{Kabashima05b}.

\subsection{Comparison between SP0 and SP$\beta$(BP)}
Now, we focus on two specific Parisi
parameters $x=0$ and $x=\beta$. A
distinctive property of these two parameter choices is that only
four out of the six variables are relevant in eqs.
(\ref{SPx_SK_simple2}) and (\ref{1RSB_Q1})-(\ref{1RSB_E}), which can
be reduced to an identical set of four equations
\begin{eqnarray}
G^{t+1}&=&J^2Q^t, \\
E^{t+1}&=&J_0m^t, \\
Q^t&=&\int Dz \tanh^2(\beta(\sqrt{G^t}z+E^t)), \\
m^t&=&\int Dz \tanh(\beta(\sqrt{G^t}z+E^t)),
\end{eqnarray}
where $G^t=\Delta^t+F^t$, $Q^t=Q_1^t$ for $x=0$ while $G^t=F^t$,
$Q^t=Q_0^t$ for $x=\beta$. As $\Delta^t$ is a non-negative variable,
this implies that $F^t$, which indicates the variance of the local
fields, for $x=\beta$ cannot be smaller than that for $x=0$ while
$E^t$, which represents the signal strength of the original message
in the local fields, is identical in both cases. Applying this
result to eq. (\ref{bit_error_rate}) leads to a conclusion that SP0
does not provide worse performance than SP$\beta$.

In terms of the decoding problem, SP$\beta$ and BP provide identical
performance. Therefore, the above argument indicates that SP0 is
never overcome by BP with respect to the error correction
performance. In order to show this, we characterise the auxiliary
distributions $P_{i \to a}^t(h)$ and $Q_{a \to i}^t(u)$ utilising
the first moments of $\tanh(\beta h)$ and $\tanh(\beta u)$,
respectively, as
\begin{eqnarray}
H_{i \to a}^t=\frac{1}{\beta}\tanh^{-1}
\left (\int dh P_{i \to a}^t(h) \tanh(\beta h) \right ),
\label{xbeta_moment1}\\
U_{a \to i}^t=\frac{1}{\beta}\tanh^{-1}
\left (\int dh Q_{a \to i}^t(u) \tanh(\beta u) \right ).
\label{xbeta_moment2}
\end{eqnarray}
Although higher moments are necessary for
completely describing general SPx
dynamics, eqs. (\ref{SPx_SK1}) and (\ref{SPx_SK2}) indicate that for
the specific choice of $x=\beta$, $H_{i \to a}^t$ and $U_{a \to
i}^t$ constitute a closed update rule as
\begin{eqnarray}
H_{i \to a}^{t+1}=\sum_{b \in {\cal N}(i)\backslash a}
U^t_{b\to i}, \label{BP1}\\
U_{a \to i}^{t}=u(H_{j \to a}^t,J_a;\beta),
\label{BP2}
\end{eqnarray}
which is nothing but the BP dynamics of finite temperature
$T=\beta^{-1}$ and are reduced to eqs. (\ref{eq:cavityupdates_rs0})
and (\ref{eq:cavityupdates_rs}) in the zero temperature limit $\beta
\to \infty$. This implies that as long as only the first moment is
concerned, which is the case for the current decoding problem,
SP$\beta$ and BP are equivalent.

\end{document}